# Streaming Multimedia over WMSNs: An Online Multipath Routing Protocol

## Samir MEDJIAH, Toufik AHMED

CNRS-LaBRI / Université de Bordeaux-1,
351, Cours de la Libération,
33405 – Talence Cedex, France.
Email: medjiah@labri.fr
Email: tad@labri.fr

## Abolghasem (Hamid) ASGARI

Thales Research & Technology (UK) Ltd.
Worton Drive, Worton Grange Business Park
Berkshire, RG2 0SB, UK
Email: hamid.asgari@thalesgroup.com

**Abstract:** Routing has become an important challenge to Wireless Multimedia Sensor Networks (WMSNs) from the standpoint of supporting multimedia applications due to the constraints on energy and computational capabilities of sensor nodes, and acquiring of the global network knowledge for disseminating packets to nodes. In this paper, we propose an online multipath routing protocol for use in WMSNs. The proposed protocol uses sensor nodes' positions to make packet-forwarding decisions at each hop. These decisions are made in real-time, in such a way that there is no need for having the knowledge about the entire network topology. This new routing protocol achieves load-balancing of traffic and minimizes energy consumption among nodes by using: (1) smart greedy forwarding scheme based on adaptive compass for selecting the most appropriate next hop for forwarding the traffic and (2) walking back forwarding scheme to bypass network holes. Performance comparison of the proposed protocols with Two-Phase Geographical Greedy Forwarding (TPGF) and Greedy Perimeter Stateless Routing (GPSR) shows that they: (a) maximize the overall network lifespan by not draining energy from some specific nodes, (b) provide quality of service delivery for video streams by influencing the best node along the route to destination, and (c) scale better in densely deployed wireless sensors network.

**Bibliographical notes**: Samir MEDJIAH received his M.S in Computer Science (Hons) from National Institute of Computer Science, Algiers, Algeria, in 2009. He is currently a PhD Student at the University of Bordeaux-1, France. His main research interests are routing, transport and congestion control of video streams in challenged networks including WMSNs, DTNs and P2P networks.

Toufik AHMED is a Professor at IPB (Institut Polytechnique de Bordeaux) in the ENSEIRB-MATMECA school of engineering. He is doing his research activities in CNRS LaBRI Lab at University of Bordeaux 1. His main research activities concern Quality of Service for multimedia wired and wireless networks, end-to-end signaling protocols, P2P network and Wireless Sensors Network.

Abolghasem (Hamid) ASGARI received his Ph.D. in Electronic Engineering from University of Wales, Swansea in 1997. He joined Thales Research & Technology (UK) in 1996 where he is a Chief Engineer. His main research interests are in QoS management and monitoring, traffic engineering, service management, and wireless sensor networking. He published about 50 papers in these areas. He served as a Visiting Professor at LaBRI in University of Bordeaux 1.

# 1 Introduction

With the advancement in miniaturization and the availability of low-cost hardware, the computing nodes embed various kinds of sensing and capturing elements including microphones and video cameras. Hence, the use of ubiquitous Wireless Multimedia Sensor Networks (WMSNs) is becoming a reality (Akyildiz et. al, 2002; Gurses and Akan, 2005; Misra et. al, 2008; Shu and Chen, 2010).

WMSNs are generally used for surveillance applications, intrusion detection, environmental and building monitoring, etc. These applications imposes additional challenges such as energy-efficient data processing both within node and in-network, audio/video bandwidth/rate adaptation to overcome the variations in networking conditions, Quality of Service (QoS) delivery to meet application specific requirements and routing and selecting appropriate paths for continual delivery of multimedia streams. Due to the distributed and dynamic nature of these types of networks, the design of a critical information infrastructure based on a WMSN raises many other challenges such as ensuring confidentiality and the integrity of the data stream, providing the means for node authentication and access control, securing routing and node grouping (Aivaloglou et. al, 2008). Among all these challenges, our work focuses on the routing and path selection issues taking into account energy constraints and QoS delivery needs.

Generally, routing in wireless sensor networks (WSN) is a challenging task. A comprehensive survey of routing protocols in WSN is given in (Al-Karaki and Kamal, 2004). A large number of research works exists to enable energy efficient routing in WSN. In fact, we can find different routing techniques that try to achieve energy-efficiency and to provide a best quality of service. One example is the multi-channel transmission in WMSNs. In (Vassis et. al, 2006), authors have evaluated the performances of routing (routing delays) when using a single and multi-channel communications in a wireless sensor and actor networks. The authors showed that the multi-channel scheme performs better than the single channel scheme especially for higher volumes of generated traffic putting the light on the important need to parallel transmissions in a wireless multimedia sensor network, where delay and packets loss are stringent constraints.

In higher layers of the communication protocols stack, performances evaluations of routing protocols for WMSNs suggests multipath routing approach to maximize the throughput of streaming multimedia traffic. This is to utilize diverse paths to route packet streams towards the destinations in order to avoid draining the energy of nodes along a specific route. In (Li *et. al*, 2010), the authors propose a multipath routing protocol based on the well known routing protocol Directed Diffusion (Intanagonwiwat *et. al*, 2000) that reinforces multiple routes with high link quality and low latency. In (Vidhyapriya and Vanathi, 2007), the authors focused on two key questions regarding multipath routing in WMSNs: (a) how many paths are needed? And (b), how to select these paths? The authors then proposed a multipath routing mechanism in order to provide a reliable transmission environment with low energy consumption by utilizing the energy availability and the received signal strength of the nodes to identify multiple routes from the source to the destination. In (Maimour, 2008), the author addresses the problem of interfering paths in a WMSN and considers both intra-session as well as inter-session interferences. The author proposes an incremental path creation mechanism where additional paths are set up only when required (typically in case of congestion or bandwidth shortage). In (Huang and Fang, 2008), authors propose MCMP (MultiConstrained MultiPath) routing protocol in order to guarantee a better QoS in terms of delay and reliability. Unlike end-to-end QoS schemes used in WSNs, the authors utilize a multiple paths creation mechanism based on local link information.
Other examples of multipath routing protocols for WMSNs include: MPMPS (*Multi-Priority Multi-Path Selection*) (Zhang et. al, 2008) and TPGF (*Two-Phase Geographical Greedy Forwarding*) (Shu *et. al*, 2008). However, these “offline multipath” protocols have to explore the multiple routes that may exist between the source and the destination before the actual data delivery phase. They may not be well adapted for large-scale highly dense network deployments and for networks with frequent node mobility.

Geographic routing is the process in which each node is aware of its geographic coordinates and uses the position of packet’s destination to perform routing decisions. These types of routing scales better for WSNs. Greedy Perimeter Stateless Routing (GPSR) (Karp and Kung, 2000) was defined as a geographic routing protocol in order for the network to scale in large size networks, i.e., to accommodate a large number of nodes having very low exchange of route state information and maintenance. The advantage of this protocol is that each node only gathers the topology information about its immediate neighbors. Thus, its greedy forwarding relies on local-knowledge for selecting the closest next hop node to the destination. This process ends up with continuous selection of the same path that leads to fast depletion of the energy of the nodes along the selected route and premature dying of these nodes.

In this paper, we examine the benefit of geographic routing along with “online” multi-path route selection process (i.e. multiple routes are created as packets advance towards the destination) and propose a new routing protocol called AGEM (Adaptive Greedy-compass Energy-aware Multipath) that takes into account both node’s energy constraints and QoS needs of audio and video streams.
The design of AGEM is driven by the following factors:

1 *Alternative paths:* multimedia applications are delay sensitive and have delay and delay variation constraints. Multimedia traffic should be delivered satisfying these requirements. In typical networks, shortest paths are heavily used for the delivery of this traffic types whereas other appropriate routers that could satisfy these traffic requirements are under-utilized.

2 *Load balancing*: In order to maximize the lifetime of WSN nodes and to avoid depletion of nodes' energy and consequently node's failures, load balancing and multi-path delivery across the network must be considered during the design of a routing protocol.

3 *Multipath transmission*: Packets in a multimedia stream are generally large in size and the transmission requirements can be several times higher than the maximum transmission capacity of sensor nodes if a single path is used for routing these packets.

4 *Online decisions*: As the topology may change from time to time, it is more appropriate to make the routing decisions in a distributed manner and in real-time. This is due to the fact that offline routing processes cannot react to topology changes and result in forwarding packets to unavailable nodes or towards disconnected routes.

5 *Node selection process*: in densely deployed networks, different neighbors may be selected as candidate for packet forwarding. To deduce an appropriate selection, the node selection process should take into account, node's energy, its distance to the destination and packet's QoS requirements.

The rest of this paper is organized as follows. Section II reviews the related work in the area of WSN routing that influenced the design of our proposed protocol. Section III presents the functionalities of proposed AGEM protocol. Section IV provides the results of performance evaluations of our proposed protocol in comparison with GPSR. Finally, section V presents our conclusions.

## 2 Related Work

In geographic routing, two greedy schemes are used to make packets progress towards the destination node. Greedy progression scheme based on distance to the destination node (Karp and Kung, 2000; Stojmenovic and Lin, 1999; Li *et. al*, 2000; Stojmenovic, 2002) and greedy progression based on angular offset in the direction towards the destination node (Kranakis *et. al*, 1999; Morin, 2001; Urrutia, 2002). In both schemes, a route between source and destination is progressively chosen only based on node-level forwarding decisions made locally at each hop.

For WMSNs, two important protocols have been proposed that make use of node positions for packet forwarding i.e., GPSR and MPMPS. MPMPS is itself based on TPGF. These protocols are further explained below.

### *2.1 The GPSR Routing Protocol*

The GPSR (*Greedy Perimeter Stateless Routing*) (Karp and Kung, 2000) was originally designed for MANETs but rapidly adapted for WSNs. The GPSR algorithm relies on the correspondence between the geographic location of nodes and the connectivity within the network by using the location position of nodes to forward a packet. Given the geographic coordinates of the destination node, the GPSR algorithm forwards a packet to destination using only one single hop location information. It assumes that each node knows its geographic location and geographic information about its direct neighbors.

This protocol uses two different packet forwarding strategies: *Greedy Forwarding* and *Perimeter Forwarding*. When a node receives a packet destined to a certain node, it chooses the closest neighbor out-of itself to that destination and forwards the packet to that node. This step is called the *Greedy Forwarding*. In case that such node cannot be found, (i.e. the node itself is the closest node to the destination out-of its neighbors but the destination cannot be reached by one hop), the *Perimeter Forwarding* will be used. The *Perimeter Forwarding* occurs when there is no neighbor closest to Destination (D) than node (A) itself. Figure 1 illustrates that node A is closer to D than its neighbors x and y. This situation is called "voids" or holes. Voids can occur due to random nodes deployment or the presence of obstacles that obstruct radio signals. To overcome this problem, *Perimeter Forwarding* is used to route packets around voids. Packets will move around the void until arriving to a node closest to the destination than the node which initiated the *Perimeter Forwarding*, after which the *Greedy Forwarding* takes over.

**Figure 1** GPSR Perimeter forwarding to bypass a void.

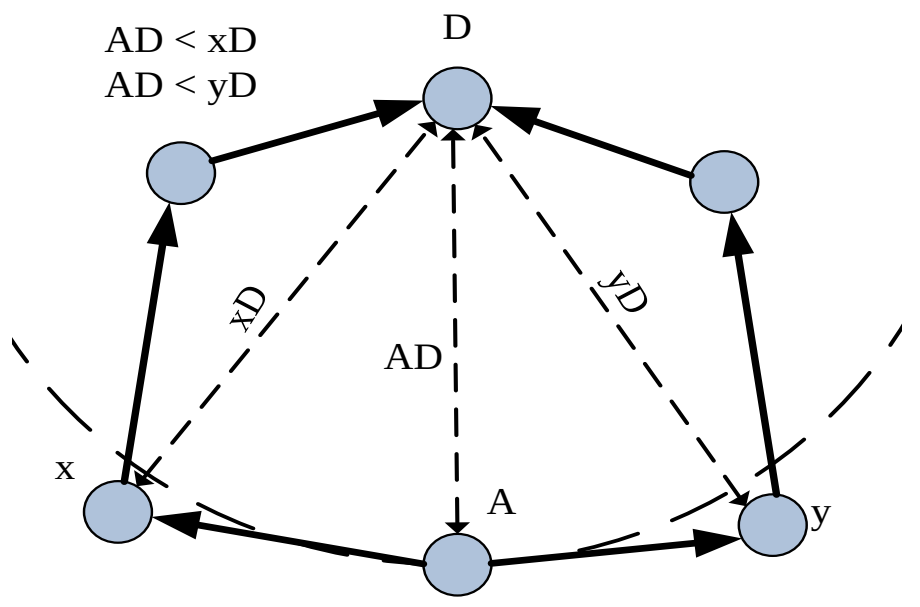

By maintaining only information on the local topology, the GPSR protocol can be suitable for WSNs. However, the greedy forwarding leads to choose only one path from the source to the destination.

## 2.2 The TPGF Routing Protocol

TPGF (*Two Phase geographical Greedy Forwarding*) (Shu *et. al*, 2008) routing protocol is the first to introduce multipath concept in wireless multimedia sensor networks (WMSNs) field. This algorithm focuses in exploring and establishing the maximum number of disjoint paths to the destination in terms of minimization of the path length, the end-to-end transmission delay and the energy consumption of the nodes. The first phase of the algorithm explores the possible paths to the destination. A path to a destination is investigated by labeling neighbors nodes until the base station. During this phase, a step back and mark is used to bypass voids and loops until successfully a sensor node finds a next-hop node which has a routing path to the base station. The second phase is responsible for optimizing the discovered routing paths with the shortest transmission distance (i.e. choosing a path with least number of hops to reach the destination). The TPGF algorithm can be executed repeatedly to look for multiple node disjoint-paths. It's worth to note that TPGF is an offline multipath routing protocol.

## 2.3 The MPMPS Routing Protocol

The MPMPS (Multi-Priority Multi-Path Selection) (Zhang *et. al*, 2008) protocol is an extension of TPGF. MPMPS highlights the fact that not every path found by TPGF can be used for transmitting video because a long routing path with long end-to-end transmission delay may not be suitable for audio/video streaming. Furthermore, because in different applications, audio and video streams play different roles and the importance level may be different, it is better to split the video stream into two streams (video/image and audio). For example, video stream is more important than audio stream in fire detection because the image reflects the event, audio stream is more important in deep ocean monitoring, while image stream during the day time and audio stream during the night time for desert monitoring. Therefore, we can give more priority to the important stream depending on the final application to guarantee the using of the suitable paths.

## 2.4 Policies for Greedy forwarding

In literature, there are different policies that can be used in geographic routing and for the selection of the next hop node. To illustrate these policies, let take '$u$' as the current forwarder node and '$d$' the destination node, then we can define these routing policies (see Figure 2):

1. ***Compass routing:*** See Figure 2(a) – The next relay node is '$v$' such that the angle $\angle vud$ is the smallest among all neighbors of '$u$' (Kranakis *et. al*, 1999).
2. ***Random compass routing:*** See Figure 2(b) – Let '$v_1$' be the node above line $(ud)$ such that $\angle v_1 ud$ is the smallest among all such neighbors of '$u$'. Similarly, define '$v_2$' to be node below line $(ud)$ that minimize the angle $\angle v_2 ud$. Then, node '$u$' randomly chooses '$v_1$' or '$v_2$' to forward the packet (Kranakis *et. al*, 1999).
3. ***Greedy routing:*** See Figure 2(c) – The next relay node is '$v$' such that the distance $\|vd\|$ is the smallest among all neighbors of '$u$' (Karp and Kung, 2000).
4. ***Most forwarding routing (MFR):*** See Figure 2(d) – The next relay node is '$v$' such that $\|v'd\|$ is the smallest among all neighbors of '$u$', where '$v$'' is the projection of '$v$' on segment $ud$ (Stojmenovic and Lin, 2001).
5. ***Nearest neighbor routing (NN)*****:** See Figure 2(e) – Given a parameter angle '$\alpha$', node '$u$' finds the nearest node '$v$' as forwarding node among all neighbors of '$u$' in a given topology such that $\angle vud \leq \alpha$.
6. ***Farthest neighbor routing (FN):*** See Figure 2(f) – Given a parameter angle '$\alpha$', node '$u$' finds the farthest node '$v$' as forwarding node among all neighbors of '$u$' in a given topology such that $\angle vud \leq \alpha$.
7. ***Greedy compass:*** Node '$u$' first finds the neighbors '$v_1$' and '$v_2$' such that '$v_1$' forms the smallest counterclockwise angle $\angle duv_1$ and '$v_2$' forms the smallest clockwise angle $\angle duv_2$ among all neighbors of '$u$' with the segment $ud$. The packet is forwarded to the node of $\{v_1, v_2\}$ with minimum distance to '$d$' (Bose and Morin, 1999; Morin, 2001).

**Figure 2** Greedy forwarding strategies: (a) Compass routing; (b) Random Compass routing; (c) Greedy routing; (d) Most Forwarding; (e) Nearest Neighbor routing; (f) Furthest Neighbor routing.

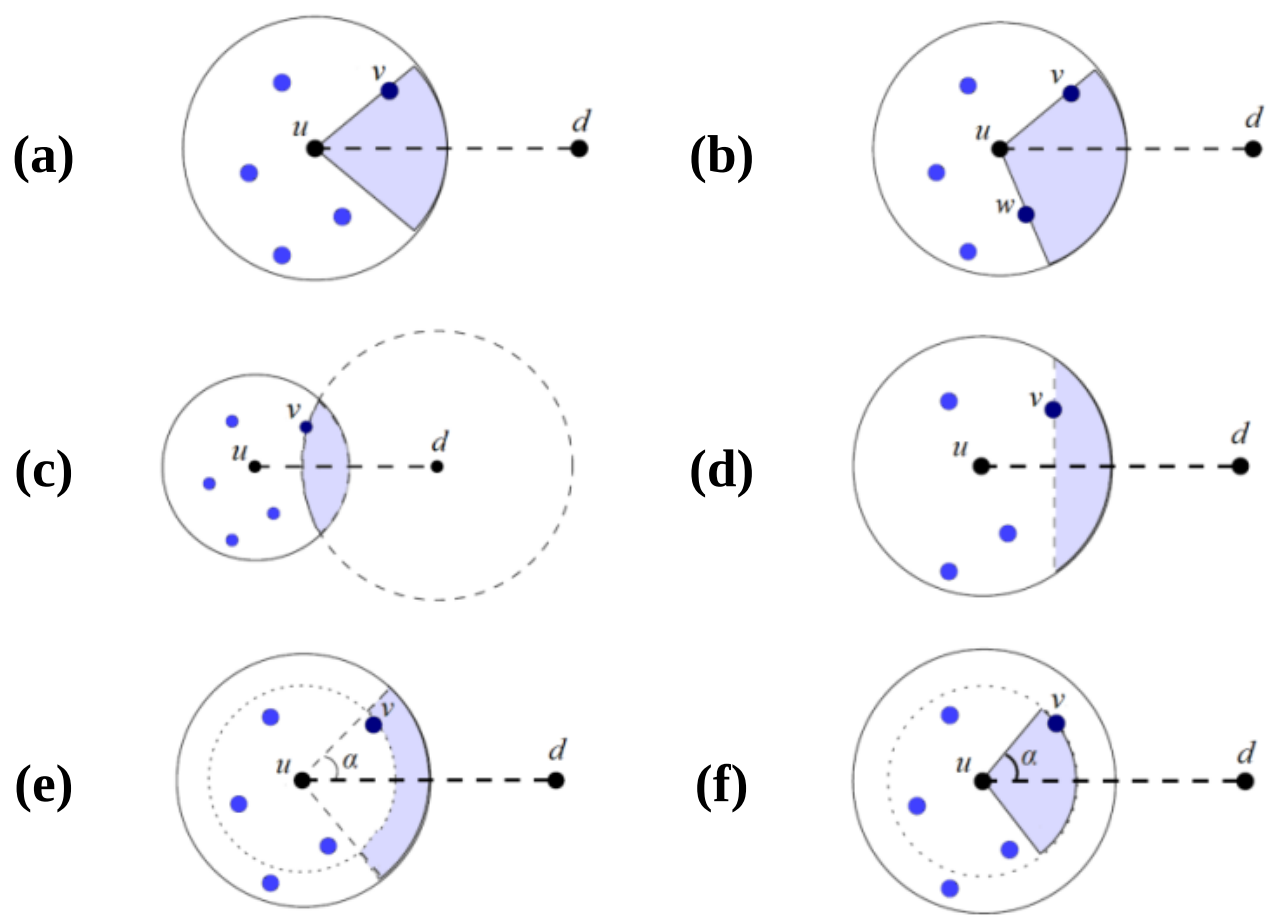

## 2.5 Discussion on routing/forwarding

Paths are selected a priori by protocols such as TPGF and MPMPS. In such cases, paths are chosen in advance from the source to the destination. Knowing the full map of the deployed network to perform routing as done by most *"offline multipath"* routing protocols is not suitable for many reasons: (1) the exchange of the network map is energy consuming, (2) the map may not reflect the current network topology, and (3) nodes' failure can be more frequent in WSN than in other ad-hoc networks. These reasons cause routing problems. In GPSR protocol, packets are forwarded hop by hop based on information available local to node i.e., the use of *"Greedy routing"* policy. GPSR seems to be more promising to scale to large network but does not achieve load balancing by making use of multiple routes.

Hence, we propose a new geographical and online routing protocol called AGEM that (1) selects neighbor nodes using an adaptive compass mechanism which is a newly defined policy, (2) routes packets on multiple paths using greedy routing policy for load balancing purposes, and (3) avoids network holes using walking back forwarding.

# 3 AGEM Routing Protocol

The main idea behind AGEM protocol is to include a load-balancing feature while being a greedy geographic routing protocol in order to increase the lifetime of the network and to reduce the queue size in the most used nodes across the network. While using a pure greedy routing protocol like GPSR, data/video streams always use the same route. In AGEM routing protocol, data/video streams are routed using different paths. At each hop, a forwarder node decides to which neighbor to send the packet. The forwarding policy at each node is based on the following four parameters: (1) the residual energy at node, (2) the number of hops visited by the packet before it arrives at this node, (3) the distance between the node and its neighbors, and (4) the history of the packets forwarded belonging to the same stream. Furthermore, only a subset of available neighbors is chosen according to the new adaptive compass selection mechanism.

The AGEM routing protocol has two modes, the *Smart Greedy Forwarding* and the *Walking Back Forwarding*. The first mode is used when there is always a neighbor node closer to the destination node than the forwarder node. The second mode is used to get out of a blocking situation in which the forwarder node can no longer forward the packet towards the destination node.

Figure **3** presents an overview diagram of AGEM routing mode switching.

The following section will explain the two routing modes.

**Figure 3** GEAMS Routing mode switching.

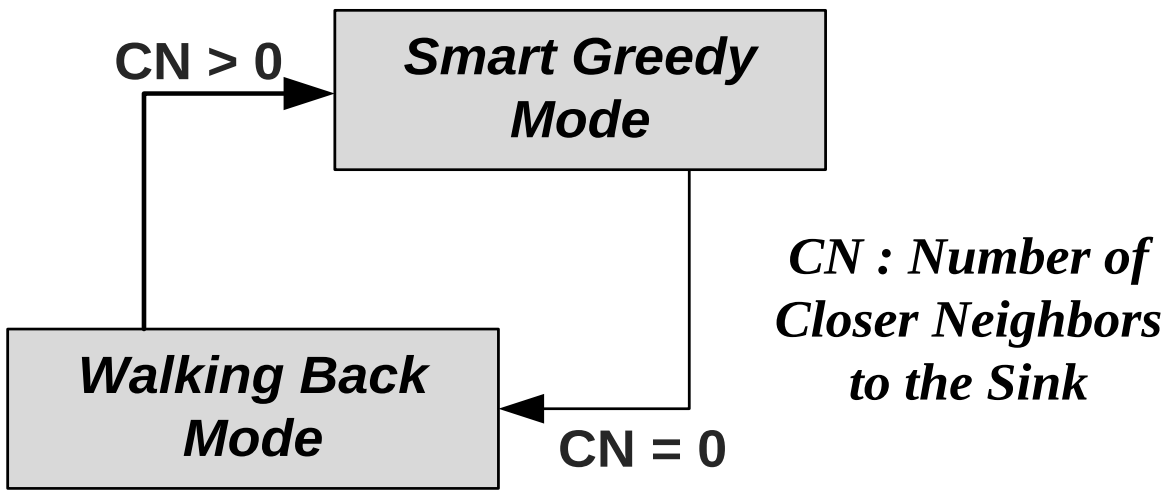

## 3.1 Smart Greedy forwarding mode:

AGEM is a geographic routing protocol where the nodes are aware of their geographic coordinates. This information can be obtained using a positioning system such as GPS or by using distributed localization techniques such as DV-Hop (Niculescu and Nath, 2003), Amorphous (Nagpal *et. al*, 2003), etc.

In AGEM routing protocol, each sensor node keeps track of related information about its *immediate* neighbors and stores the information that includes the estimated distance to its neighbors, the distance of the neighbor to the destination, the data-rate of the links, and the remaining energy of neighbors. This information is updated by the mean of beacon messages propagated locally, scheduled at fixed adjustable intervals. Relying on this information, a forwarder node will give a score to each neighbor according to a function (i.e. "*f(x)*").

Since AGEM protocol is an online protocol and relies on beacon exchange for neighborhood state maintenance, AGEM can be used for static or mobile sensor networks.
Since AGEM routing algorithm is based on geographic coordinates, distance-based greedy progression is used along angle-based greedy progression for next hop node selection. So, not all the neighbors closest to the destination than the forwarder node are going to be selected as the candidates for packet forwarding. This set of nodes is reduced to only include those nodes with best angular offset towards the destination.

At the beginning, the forwarder node chooses only neighbor nodes that are within an angular (α) view towards the destination with an initial angle of $\alpha_0$ (e.g., $\alpha = \alpha_0 < 30°$). A minimum of *"n"* neighbor nodes (neighboring set with n>=2) must be found to perform load balancing. If *n*=1 then there is just one node set where no load balancing can be achieved. If no node is found, the angle α is incremented by **Δ**α (e.g., **Δ**α =10°) until it reaches 180°. At this stage, if no node is found then a walking back forwarding is needed since the forwarder is facing a hole. Figure 4 illustrates this adaptive forwarding policy.

**Figure 4** AGEM adaptive compass policy.

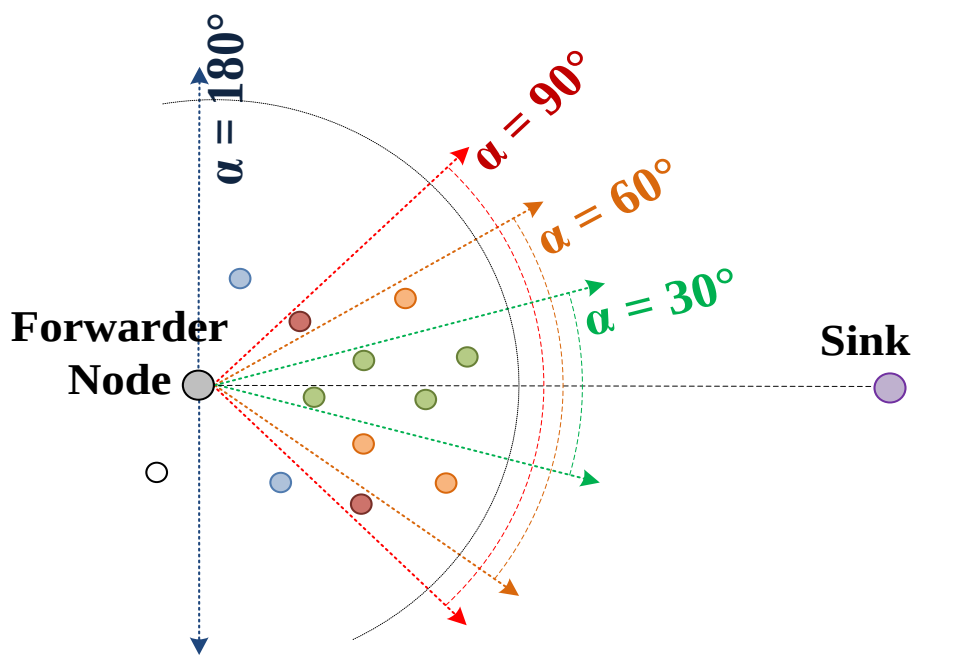

Choosing a node from the neighboring set to forward a packet will depend on the score given to each node according to the "*f(x)*" function (see Figure 5). The *f(x)* considers the energy consumption which is defined in the following subsection.

**Figure 5** One-hop neighbors sorted according to their scores.

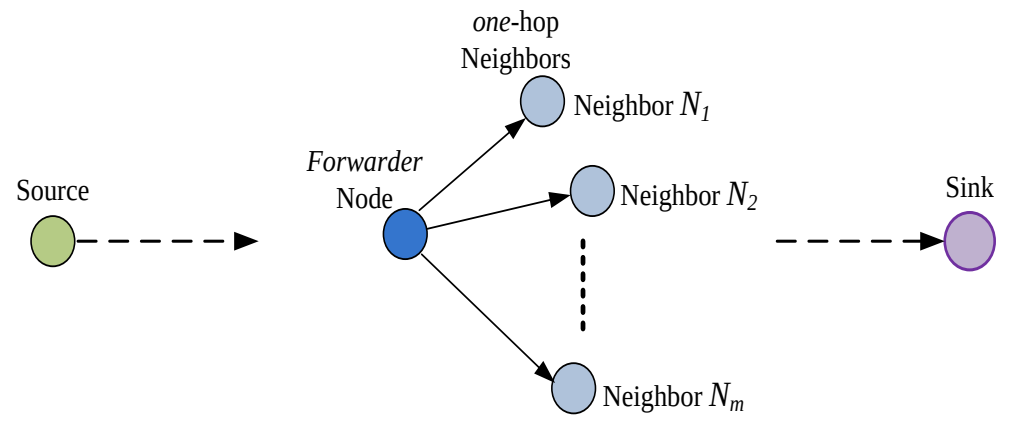

*Packet energy consumption :*

When a node ($A$) sends a packet ($pk$) of $n$ bits size to a node ($B$), the energy of node ($A$) will decrease by $E_{TX}(n, \overline{AB})$ while the energy of the node $B$ will decrease by $E_{RX}(n)$. Consequently, the cost of this routing decision is $E_{TX}(n, \overline{AB}) + E_{RX}(n)$ considering the energy of the whole network. Figure 6 illustrates this energy consumption.

**Figure 6** Packet energy consumption considering two communicating nodes A and B.

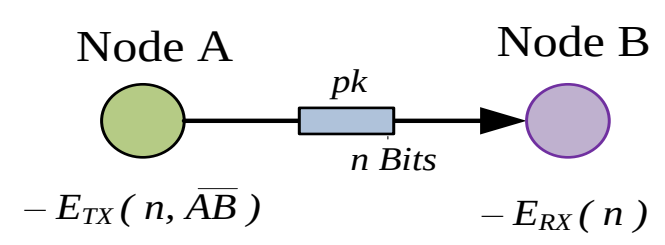

We assume that the transmitted data packets in the network have the same size. We propose an objective function to evaluate a neighbor $N_i$ for packet forwarding. This objective function takes into account the packet energy consumption and also the initial energy of that neighbor. The proposed objective function can simply be:

$$f(N_i) = N_{i_{Energy}} - E_{TX}(N_{i_{Distance}}) - E_{RX}$$

Where: $E_{TX}(D)$ is the estimated energy to transmit a data packet through a distance D, and $E_{RX}$ is the estimated energy to receive the data packet.

These two functions rely on the energy consumption model proposed by (Heinzelman *et al.*, 2000). According to this model, we have:

$$E_{TX}(k, D) = k \cdot (E_{ELEC} + \varepsilon_{amp} \cdot D^2)$$

$$E_{RX}(k) = k \cdot E_{ELEC}$$

Where:$\boldsymbol{k}$ is the size of the data packet in bits, $\boldsymbol{D}$ is the transmission distance in meters, $\boldsymbol{E_{ELEC}}$ is the energy consumed by the transceiver electronics,$\boldsymbol{\varepsilon_{amp}}$ is the energy consumed by the transmitter amplifier.$E_{ELEC}$ was taken to be 5 $\mu J/bit$ and $\varepsilon_{amp}$ 1 $\eta J/bit$.

Upon receiving a data packet from the source node $s_i$, the forwarder node retransmits the packet to a neighbor that is closest to the destination node and in such a way that the number of hops the packet traversed, will meet the rank of that neighbor (neighbors are ranked according to their score). The main idea is to forward a packet with the biggest number of hops through the best neighbor, and consequently a packet with the smallest number of hops is routed through the worst neighbor to allow a proper load balancing in the network (see Figure 8 and Figure 9). Figure 7 describes an algorithm as the forwarding policy.

For each known source node $s_i$ a forwarder node (N) maintains a pair ($H_i$, j). $H_i$ represents the mean hop count that separates $s_i$ from *N*, and *j* represents the neighbor (Nj) whom score (i.e. *f(x)* function) is closest to the average score of all closest nodes to the sink in the neighbor set (called best neighbor set).

**Figure 7** The Smart Greedy Forwarding algorithm.

***Upon_Recieving_a_Packet*** ( *pk* )

<u>*Parameters:*</u>

***Best_Neighbor***: *a set of the closest neighbors to the sink node sorted in descending order by their score {$BN_1$, $BN_2$, … $BN_m$}.*

***m*** = |***Best_Neighbor***|. ***m*** represents the cardinal of the *Best_Neighbor set*

***j*** *:index of the node in the set* ***Best_Neighbor*** *whom score is closest to the average score of all closest nodes to the sink. For example, if* ***Best_Neighbor is {8,5,2,1}*** *the average score is* ***4*** *then* ***j=2*** *(starting from index=1)*

<u>*Functions:*</u>

***Get_Hop_Values ($S_i$)*** *returns the stored values of empirical hop count from already known source $S_i$ and the j index of the average score of all closest nodes to the sink. These values are ($H_i$, j)*

***Set_Hop_Values ($S_i$, $H_i$, j)*** *sets the empirical hop count for source $S_i$ to be $H_i$ and j to be the index of the average score of Best_Neighbor set.*

***Forward (pk, $BN_k$ )*** *forwards the packet* ***pk*** *to the neighbor* ***k*** *which has $BN_k$ score*

```
01: if (Get_Hop_Values (pk.SourceNode) is Null ) {
02:     Forward (pk, BN1)          // Default forward to best node
03:     H ← pk.HopCount
04:     Set_ Hop_Values (pk.SourceNode, H, j)
05:  }
06: else {      //Get_Hop_Values (pk.SourceNode) is not null
07:     (H,j) ← Get_Hop_Values (pk.SourceNode)
08:     Δh ← H – pk.HopCount
09:     index ← j + Δh
10:     case (index  ≤  0) {
11:         H ← H–index +1
12:         index←1  // index of the best node in neighbor_Set
13:     }
14:     case ( index > m )  {
15:         H ← H–index+m
16:         Index ←m //index of the worst node in neighbor_Set
17:     }
18:     Forward ( pk, BNindex ) // Smart forward
19:     Set_ Hop_Values ( pk.SourceNode, H,j)
20: }
21:
```

As shown in Figure 7, the algorithm checks (Line 1) if a packet is already received from a source node. If no, the packet will be always forwarded to the best node (line 2), and the hop count "H" and the average score index "j" in the best neighbor set are set. These empirical values will be used later to allow load balancing. It is clear that the first packet received from an unknown source will be always forwarded to the best neighbor node.

Line 7 specifies that we have already an empirical estimation of the hop count H and the average index *j* from a particular source. These values are retrieved as shown in line 8. We calculate (in line 9) the deviation Δh of the hop count of the received packet compared to the stored value H. The index of the new forwarder neighbor that allows best load balancing will be adjusted by Δh (line 10). However, two different out of range situations may occur. Line 11 specifies that the received packet has passed through a lot of hops, and thus it needs to be forwarded to the best node (i.e. node with index=1). The received packet that has experienced a less hop count than the empirical value H (line 15), and thus it has to be forwarded to node with higher index (index=m). The new empirical value is computed (Line 12 and 16) that will be used later as a new reference. Finally, the packet is forwarded by using the described *Smart Greedy Forwarding* (line 19).

## 3.2 *Walking Back forwarding mode*

Because of node failures, node energy depletion due to processing and scheduling activities and node mobility, disconnections may occur in a WSN generating what we call "voids". At certain times, a forwarder node may face a void where there is no closest neighbor to the sink as illustrated in Figure 10.

**Figure 8** Forwarding the first packet of a data stream.

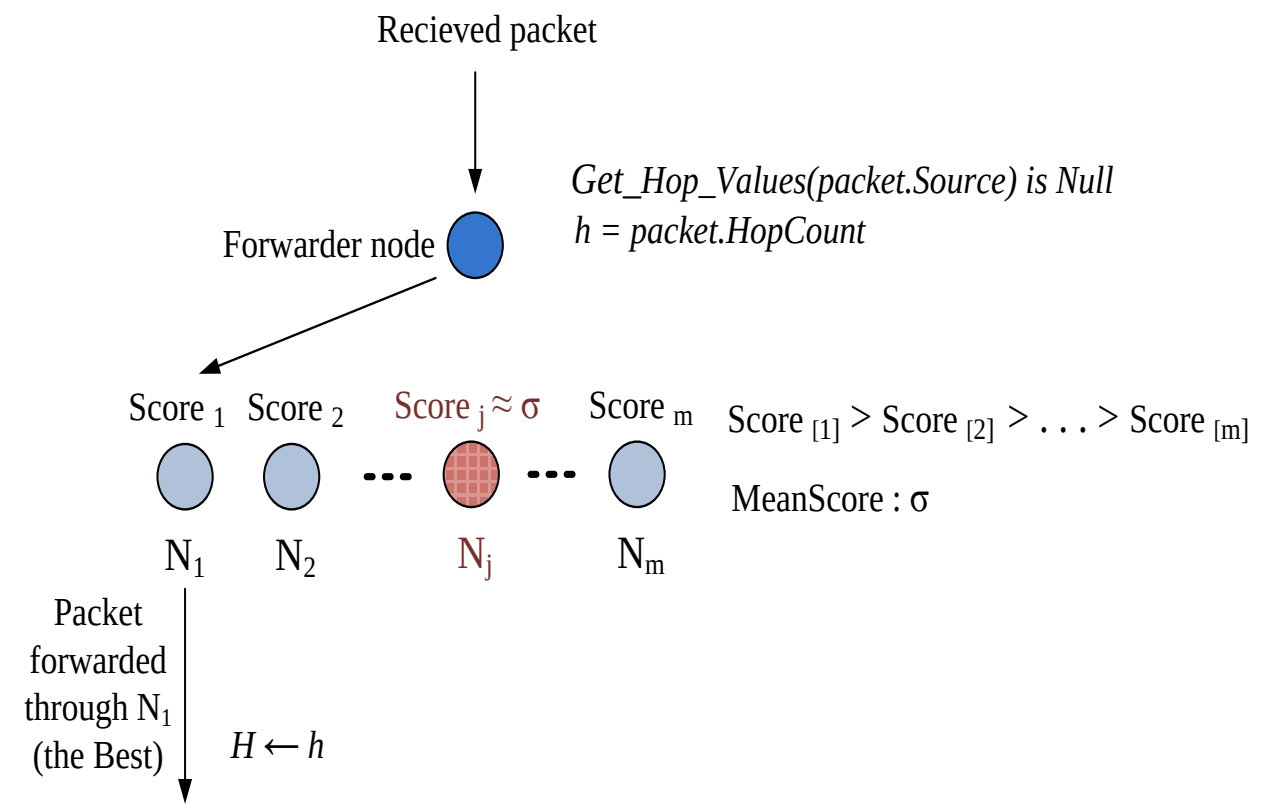

**Figure 9** Forwarding a packet of an already known data stream.

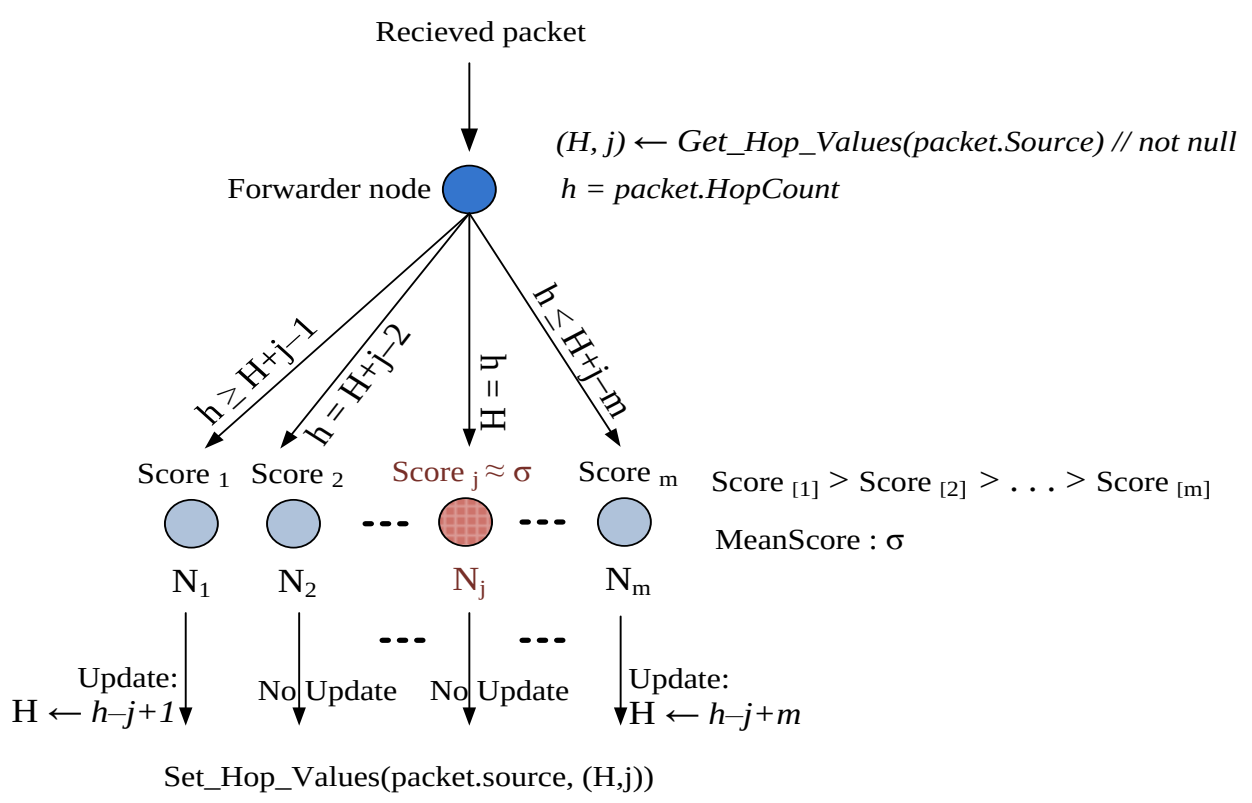

In this case, the node enters the walking back forwarding mode in order to bypass this void. In such a case (see Figure 10), the forwarder node will inform all its neighbors that it cannot be considered as a neighbor to forward packets to the sink. This node will also delegate the forwarding responsibility to its nearest neighbor to bypass the void. This process does recursively step back until a node is found that can forward the packet successfully.

**Figure 10** A blocking situation where a forwarder node has no neighbor closer to the sink than itself.

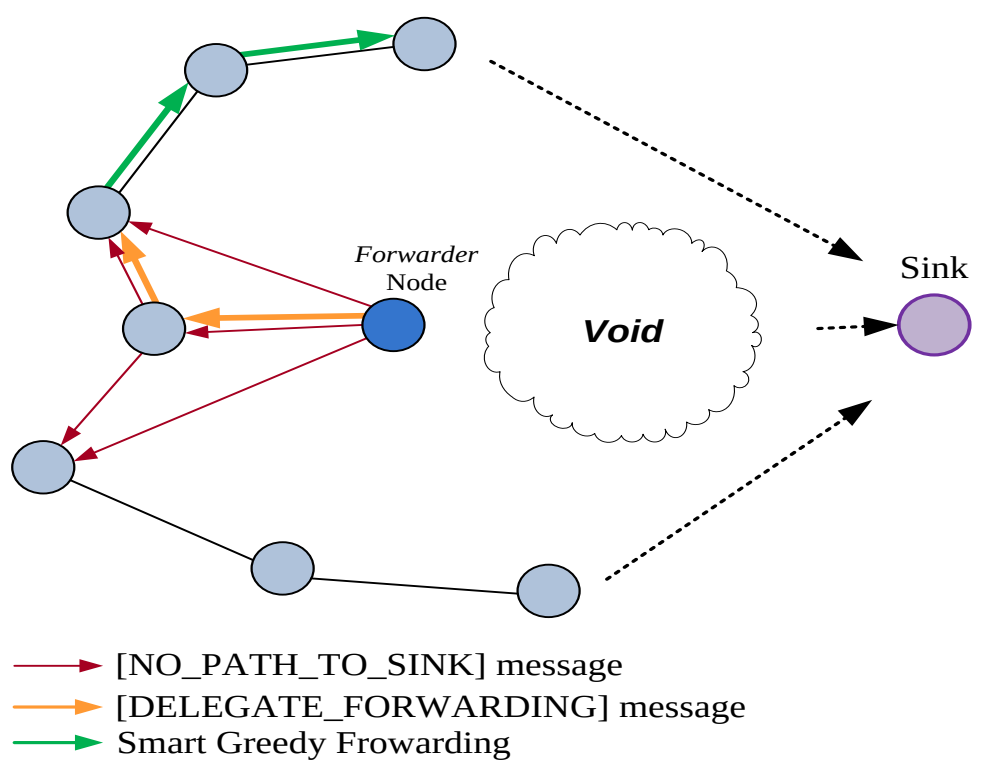

This technique is better than the perimeter routing mode used in GPSR, since this kind of process is only done once a packet is received from an unknown stream, all the other packets belonging to the same stream will be routed avoiding the nodes that are facing a void toward the sink.

## 4 Simulation and Evaluation

### *4.1 Simulation Environment*

We have considered a homogenous WMSN, in which, nodes are randomly deployed through the sensing field. The sensing field is a rectangular area of 500m x 200m. The sink node is situated at a fixed point in the righter edge of the sensing field at coordinates (490, 90) while a source node is placed in the other edge at coordinates (10, 90). We have considered this network for video surveillance (see Figure 11). In response to an event, the source node will send images with a rate of 1 image per second during 30 seconds.

**Figure 11** Data Delivery in Response to an Event in a WMSN.

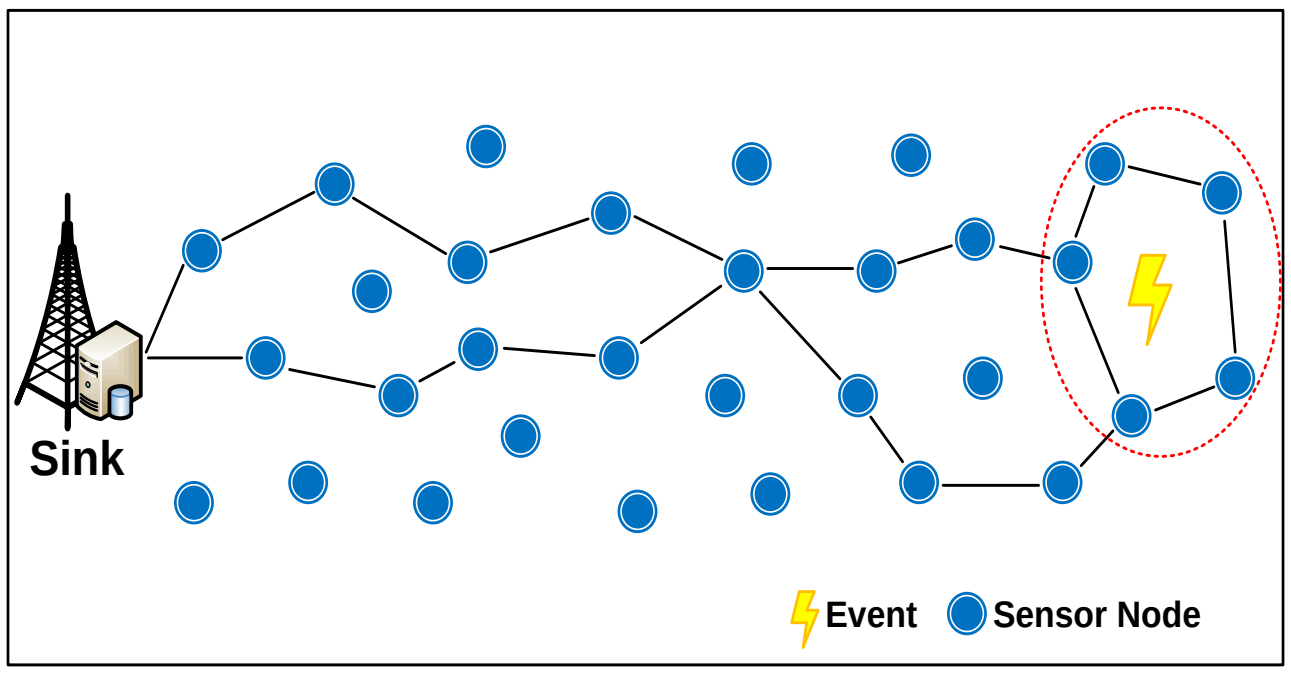

To demonstrate and evaluate the performance of our proposed protocol AGEM, we used OMNeT++ 4 which is a discrete event network simulator (Varga and Hornig, 2008). To prove the effectiveness of AGEM, we have also implemented the GPSR algorithm (as an online but single-path routing protocol) and an adapted version of MPMPS on top of the TPGF algorithm (as an offline-multipath routing protocol) and we compared the simulation results. We have also introduced GEAMS (Greedy Energy-Aware Multipath Stream-based) Routing protocol which consists of a "light" version of AGEM that does not include the adaptive compass mechanism for next hop node selection. Thus, GEAMS uses only distance-based greedy progression. Table 1 summarizes the simulation environment. We have considered that the link data is of type IEEE 802.15.4.

**Table 1** Simulation parameters.

| Parameter | Value |
|---|---|
| Network Size | 500m x 200m |
| Number of Sink Nodes | 1 |
| Number of Source Nodes | 1 |
| Number of Sensor Nodes | 30, 50, 80 |
| Number of Images | 30 images |
| Image Size | 10Kb |
| Image Rate | 1 image/sec |
| Maximum Radio Range | 80 meters |

To evaluate the performance of our protocol, we have considered the following three topology types:

#### *4.1.1 Plain topology:*

This topology is used to evaluate the behavior of the routing algorithm especially the smart greedy forwarding mode.
Here, we have used three plain topologies; a network of 30, 50 and 80 sensor nodes. An example of these topologies is shown in Figure 12.

**Figure 12** A 30-nodes network topology.

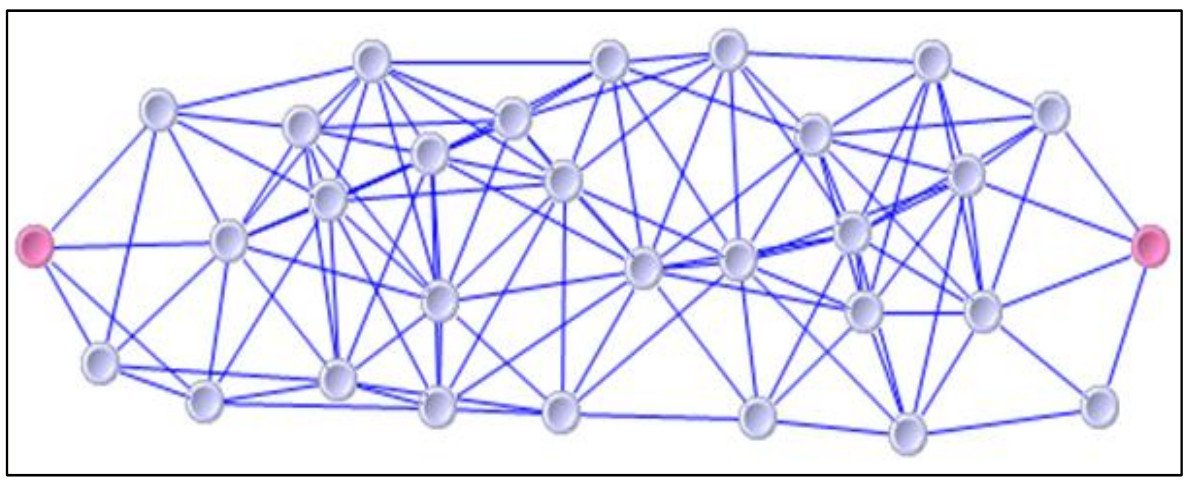

#### *4.1.2 Topology with holes:*

This topology is used to evaluate the performance of the routing algorithm in presence of holes (i.e. to evaluate the performance of the walking back forwarding mode).
We have used four topologies with holes; a network of 30 sensor nodes with one or two holes, and a network of 50 sensor nodes with one or two holes. An example of such topologies is shown in Figure 13.

**Figure 13** A 30-nodes network topology with two holes.

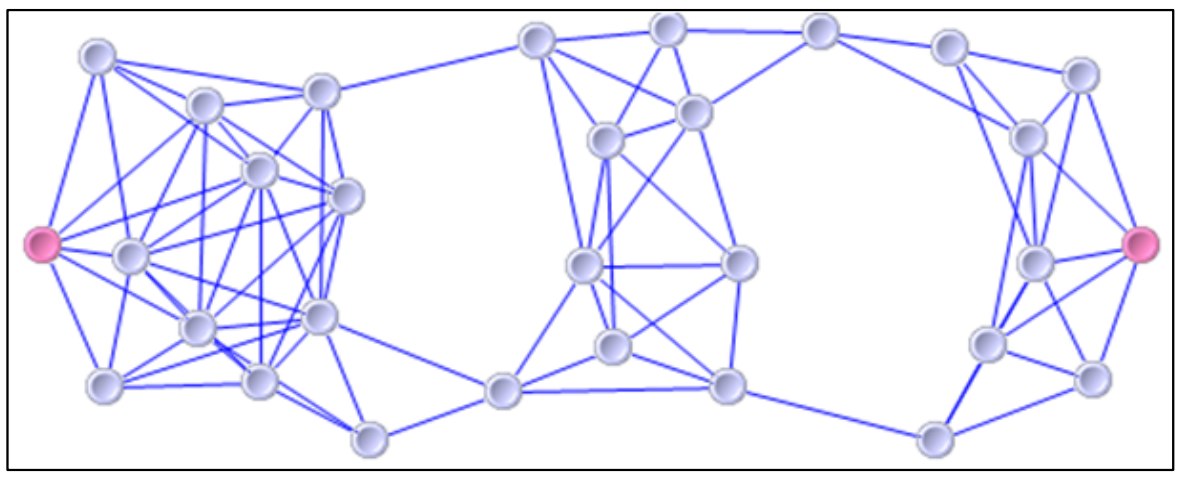

### *4.1.3 Regular topology:*

This topology is used to evaluate the load-balancing feature of the algorithm. We have used one grid topology of 26 sensor nodes. This network is shown in Figure 14.

**Figure 14** A 26-nodes grid network topology.

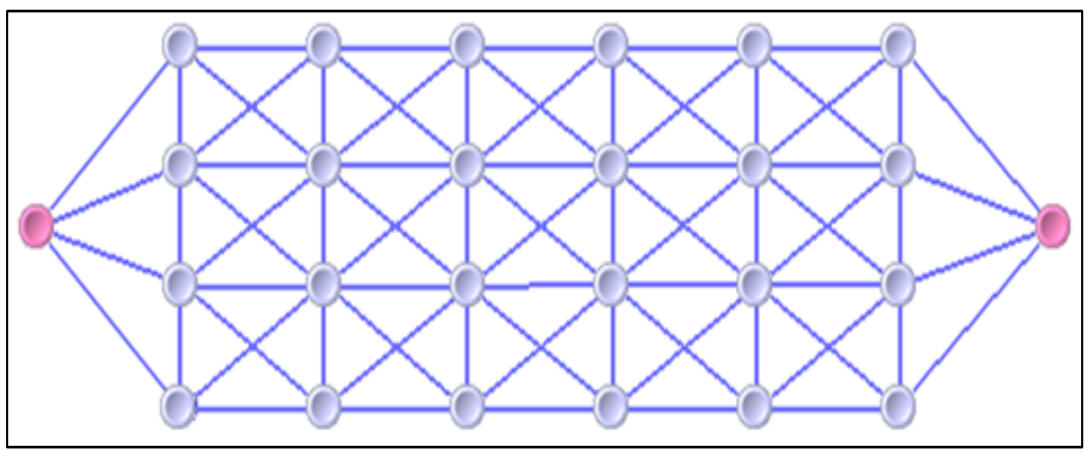

In all of the above topologies, we consider the minimum distance between two neighboring nodes to be greater than 1 meter. For each topology, we have measured various metrics:

1. ***Global Energy Distribution (GED)***: it is the average and the standard-deviation of the residual energy at all network nodes.
2. ***Local Energy Distribution (LED)***: it is the average residual energy in contiguous regions of 40 meters width.
3. ***End–to–End Delay Distribution***: it is the average and the standard-deviation of the end-to-end delay.
4. ***Packet Loss Ratio***: it is the percentage of lost packets during the transmission.

## *4.2 Simulation Results:*

In this section, we only present the simulation results obtained for different topologies using GPSR, TPGF, GEAMS and AGEM. The next section provides the discussion on the results obtained:

### *4.2.1 Plain topologies*

The distribution of the residual energy in the network (GED) is shown in Figure 15.

**Figure 15** Average residual energy in “plain” topologies.

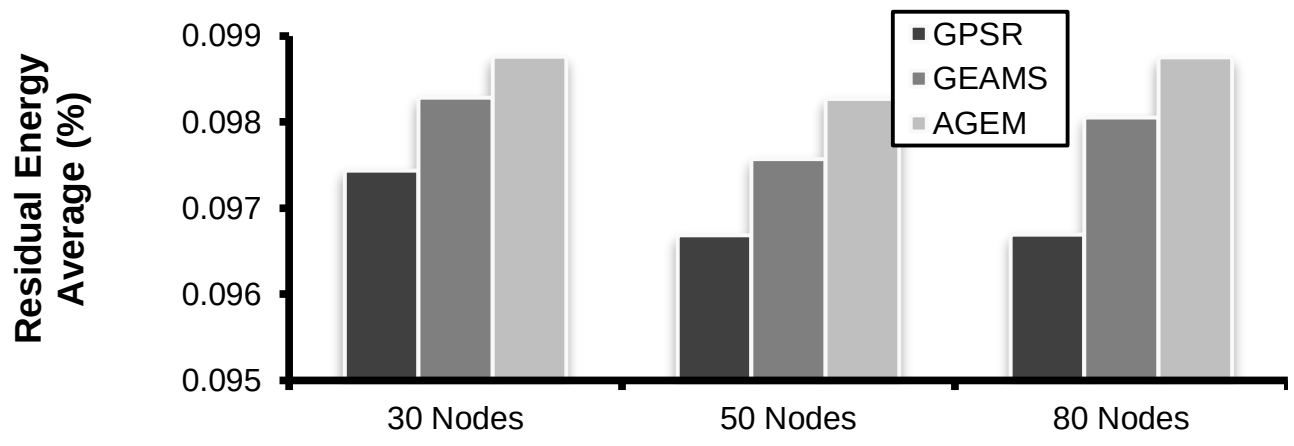

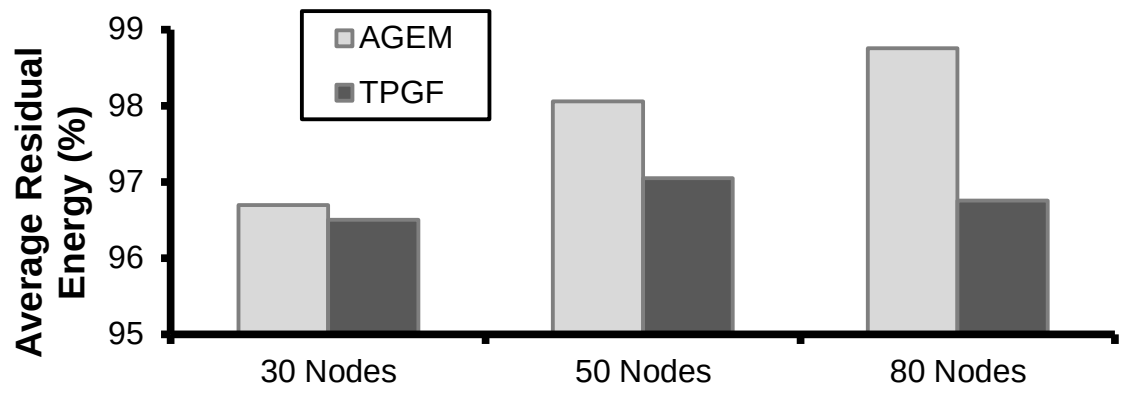

The distribution of the residual energy across the network (LED) is shown in figures 16-18.

**Figure 16** The distribution of the residual energy across the network for 30-node network topology.

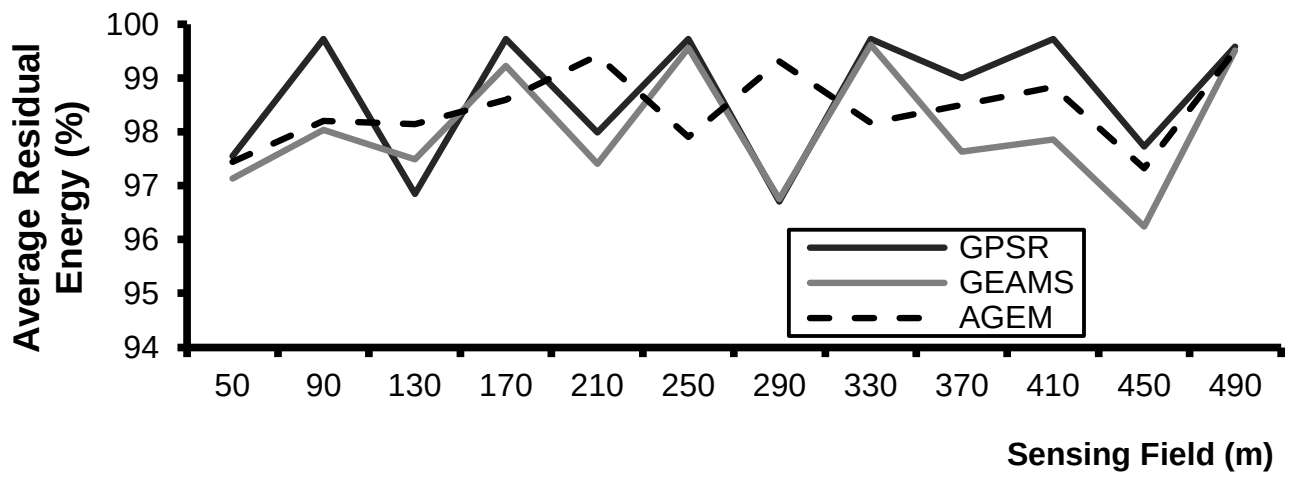

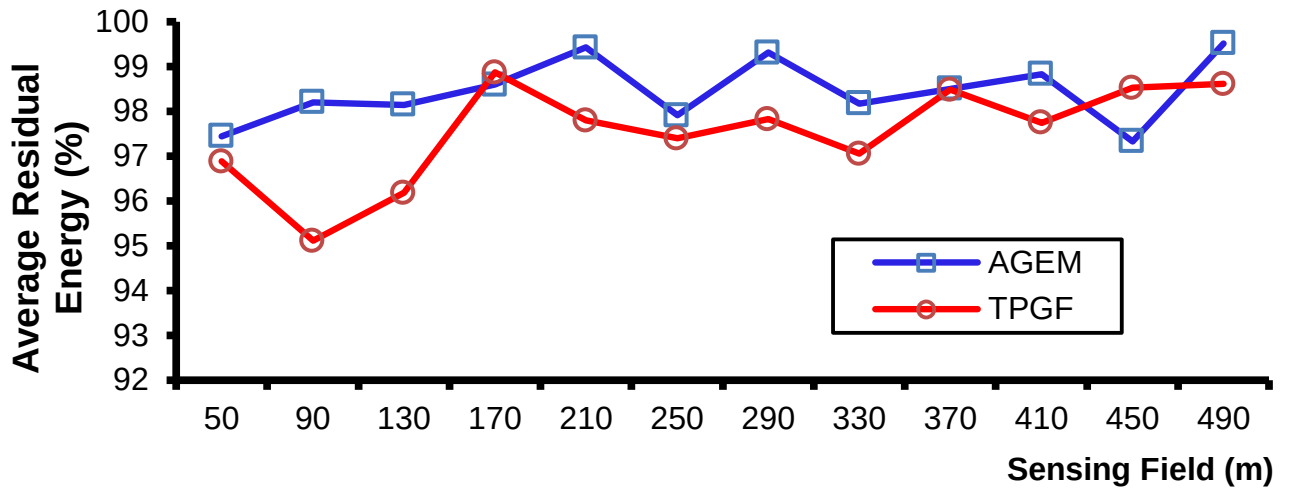

**Figure 17** The distribution of the residual energy across the network for 50-node network topology.

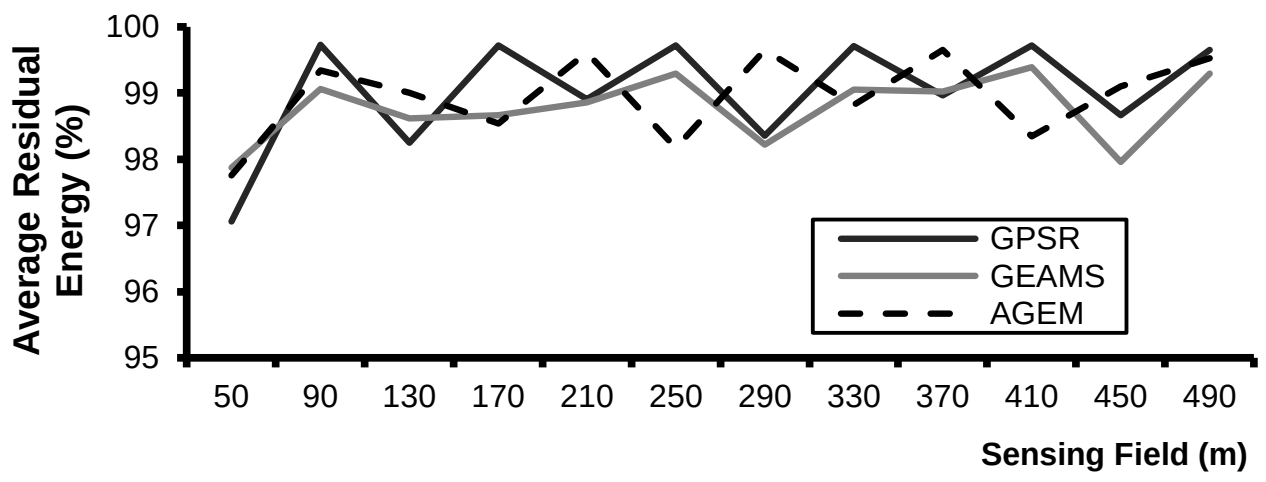

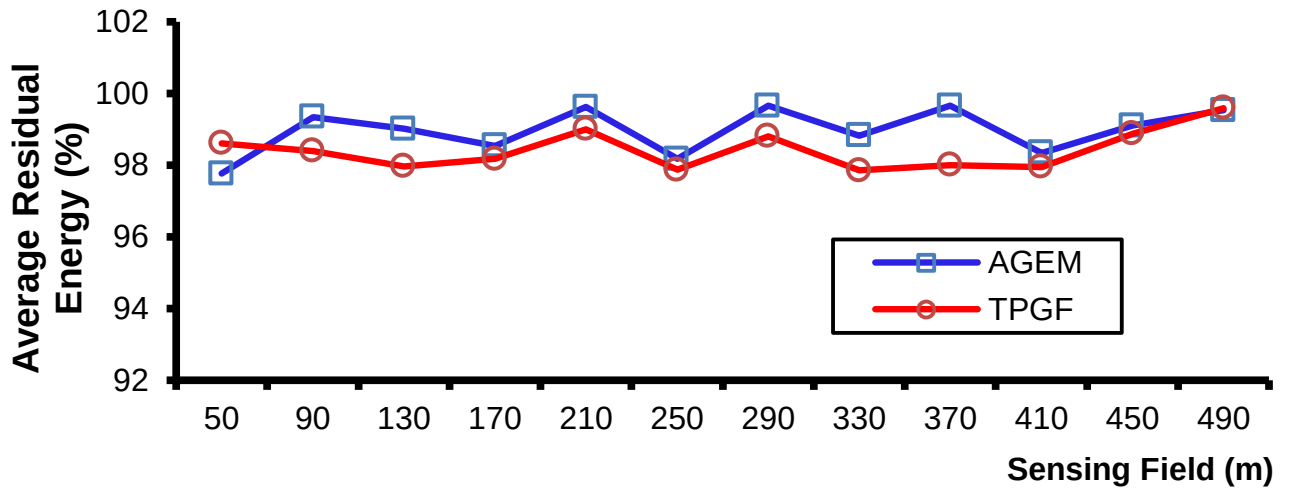

**Figure 18** The distribution of the residual energy across the network for 80 nodes topology.

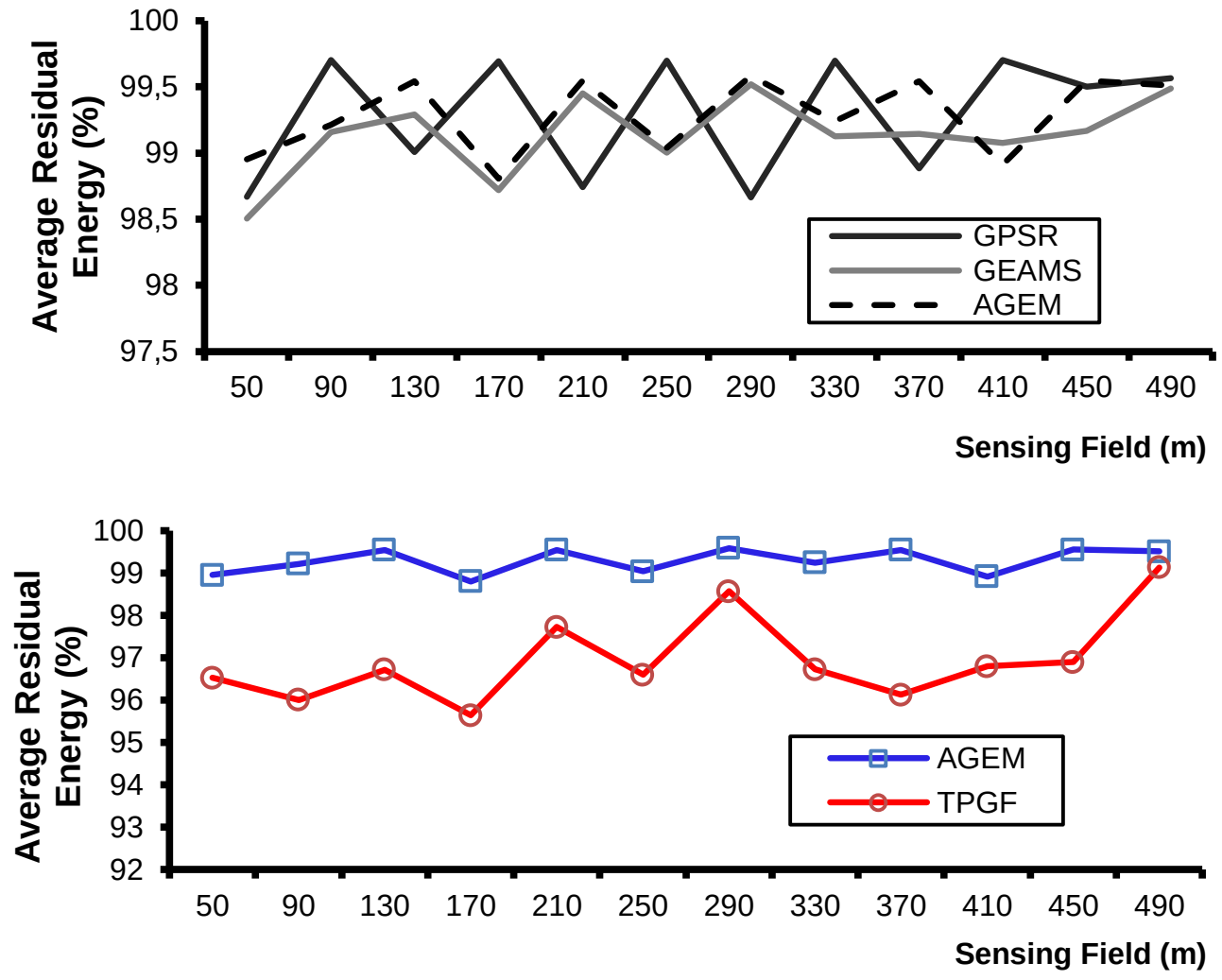

The distribution of the end-to-end delay is shown in Figure 19.

**Figure 19** Average end-to-end delay in plain topologies.

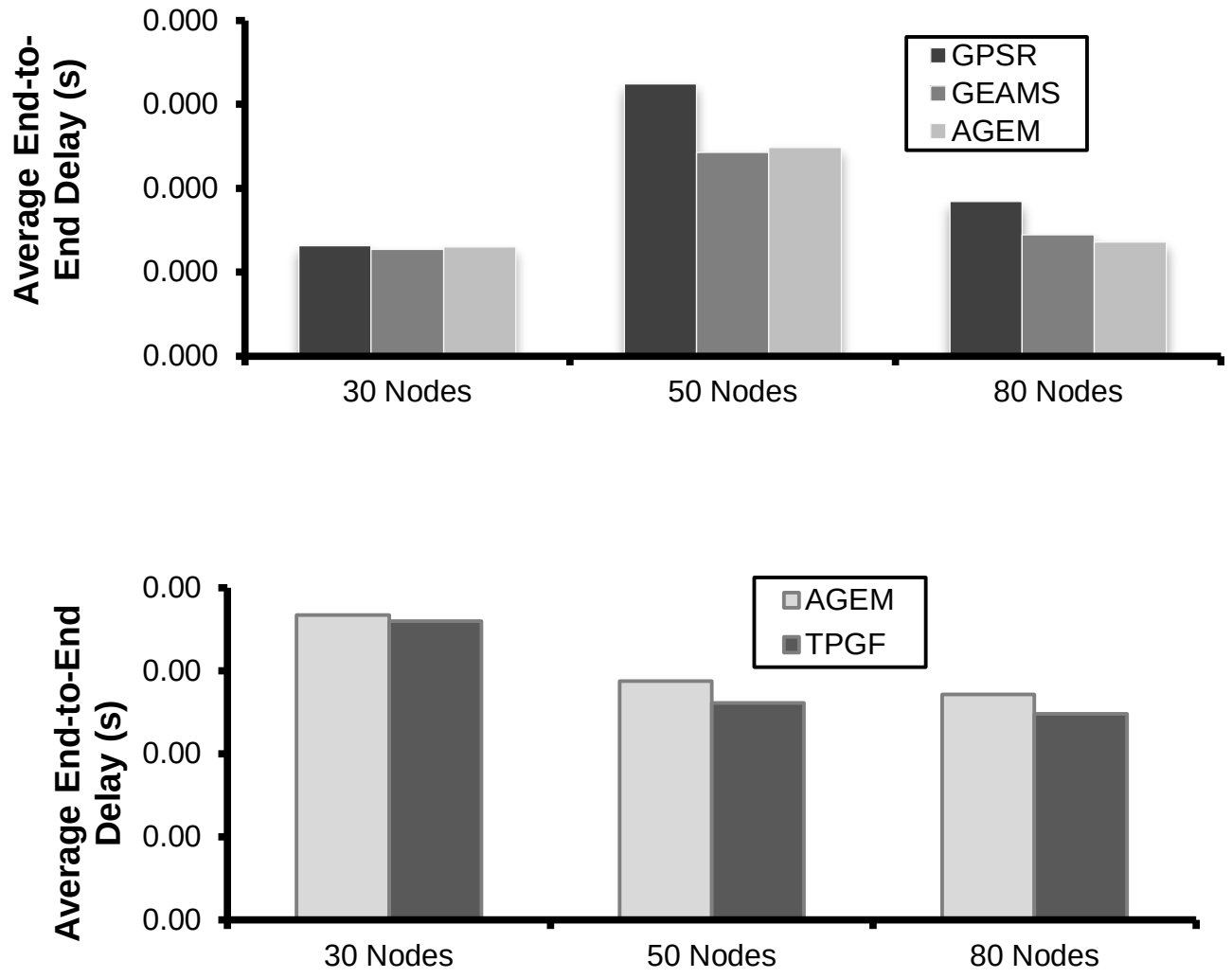

The packets loss ratio during image transmission is shown in Figure 20.

**Figure 20** Packet-loss ratio in plain topologies.

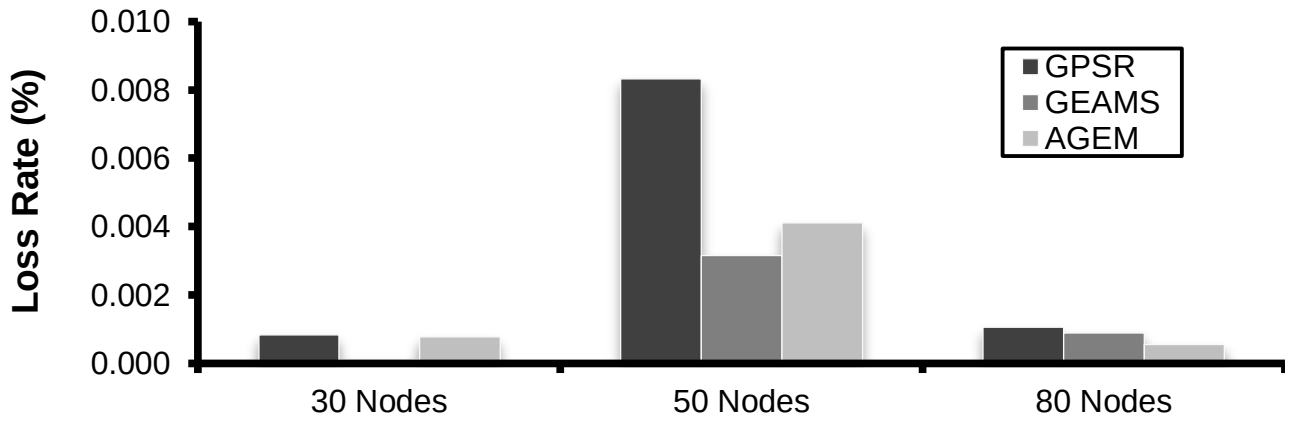

### *4.2.2 Topologies with holes*

The distribution of the residual energy in the network (GED) is shown in Figure 21.

**Figure 21** Average residual energy in topologies with holes.

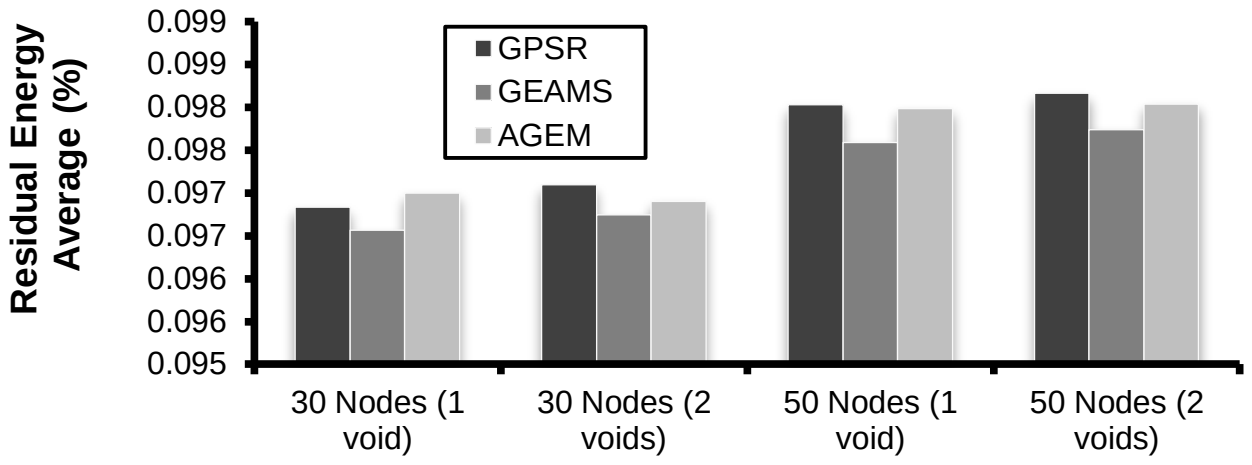

The distribution of the residual energy across the network (LED) in a topology with holes is shown in Figure 22.

**Figure 22** Residual energy distribution across the network for 50-node network topology with two holes (holes are in region 210m-290m along the sensing field)

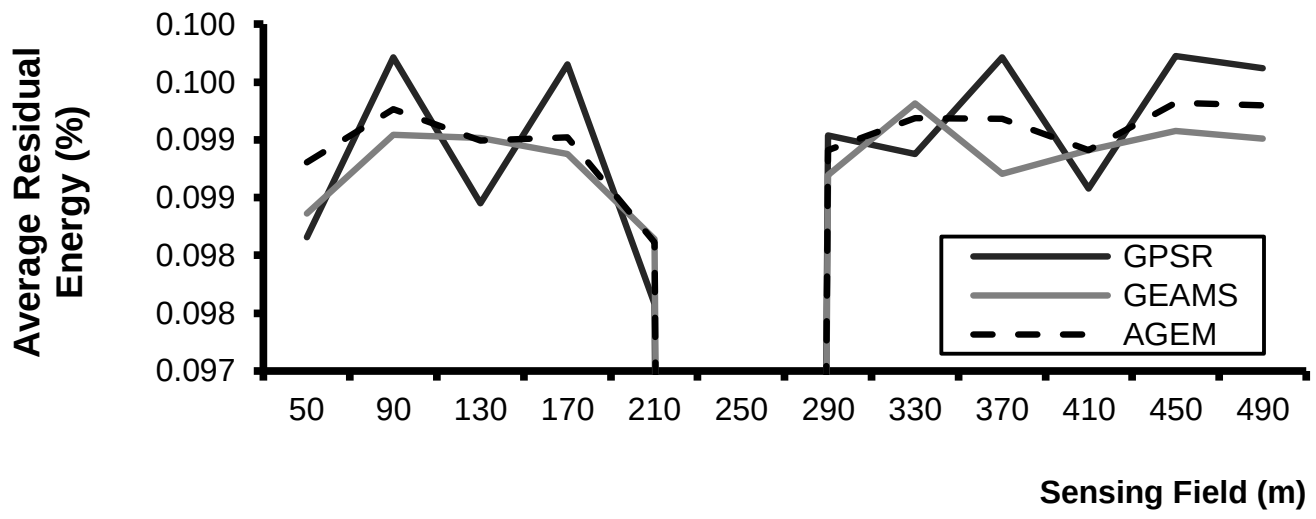

The distribution of the E2E delay is shown in Figure 23.

**Figure 23** Average end-to-end delay in topologies with holes.

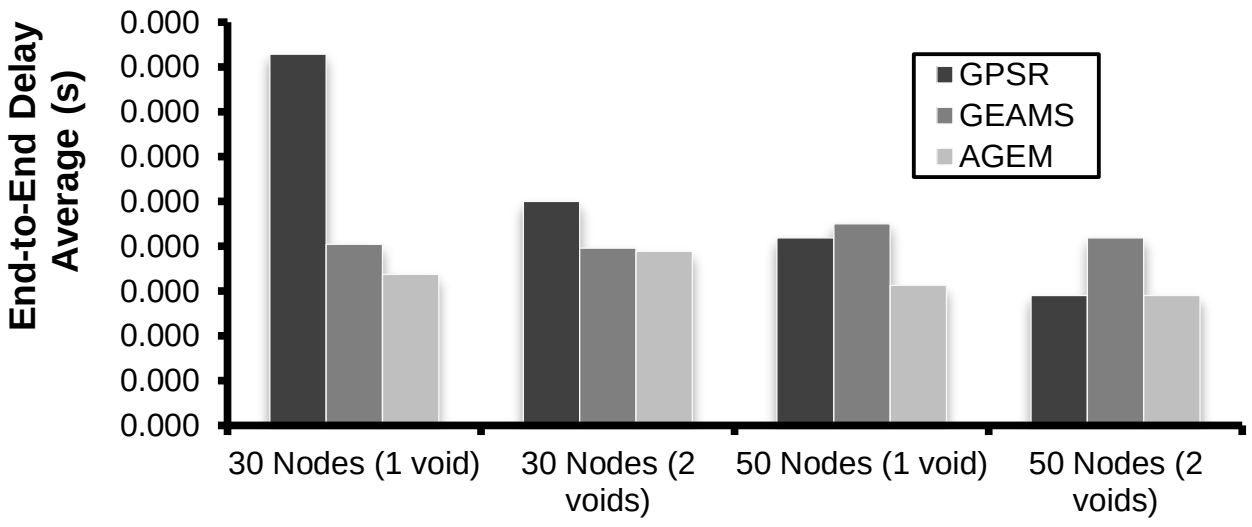

The ratio of overall packet losses during the transmission is shown in Figure 24.

**Figure 24** The packet-loss ratio in topologies with holes (please note the logarithmic scale)

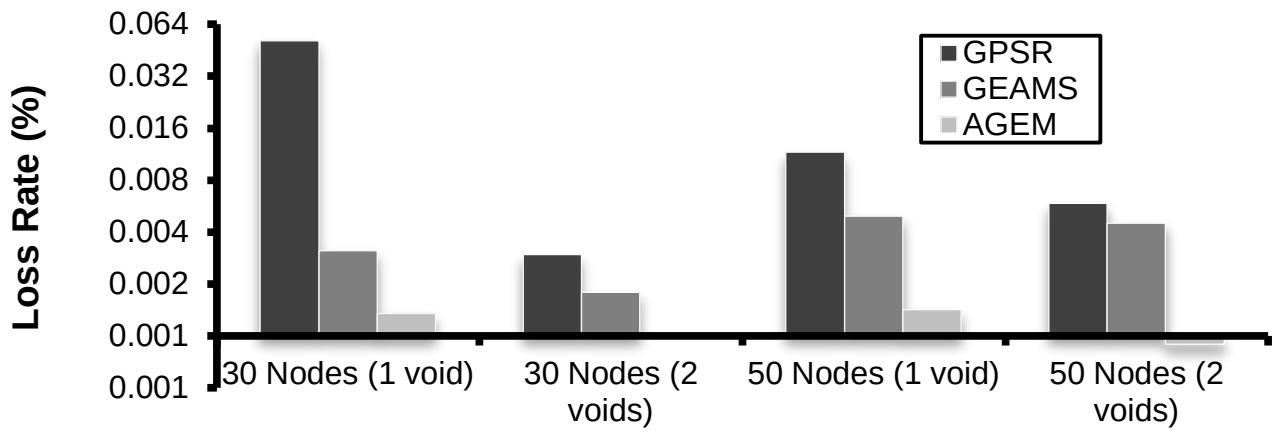

### 4.2.3 *Regular topology*

To illustrate the load-balancing feature of AGEM, we have used a grid topology and simulated a transmission between nodes *Src* and *Dest* as shown in figures 25-26. The figures show the residual energy at each node by the mean of a graduated color that corresponds to their residual energy (Red to 0% and Blue to 100%).

**Figure 25** Residual energy with GPSR in a grid topology.

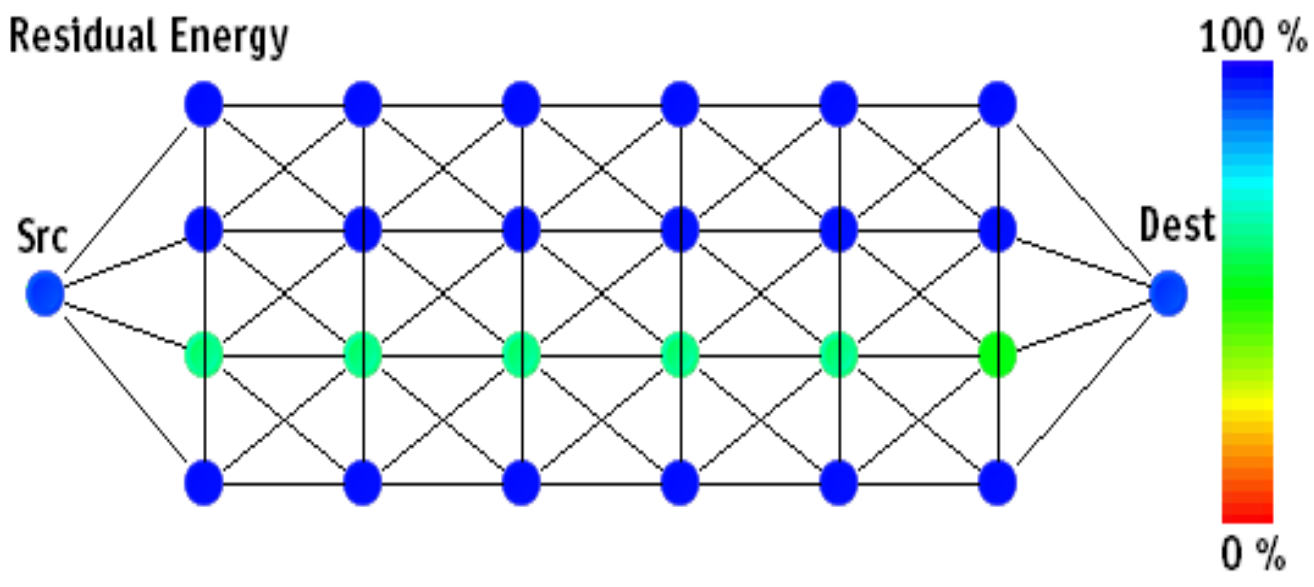

**Figure 26** Residual energy with AGEM in a grid topology.

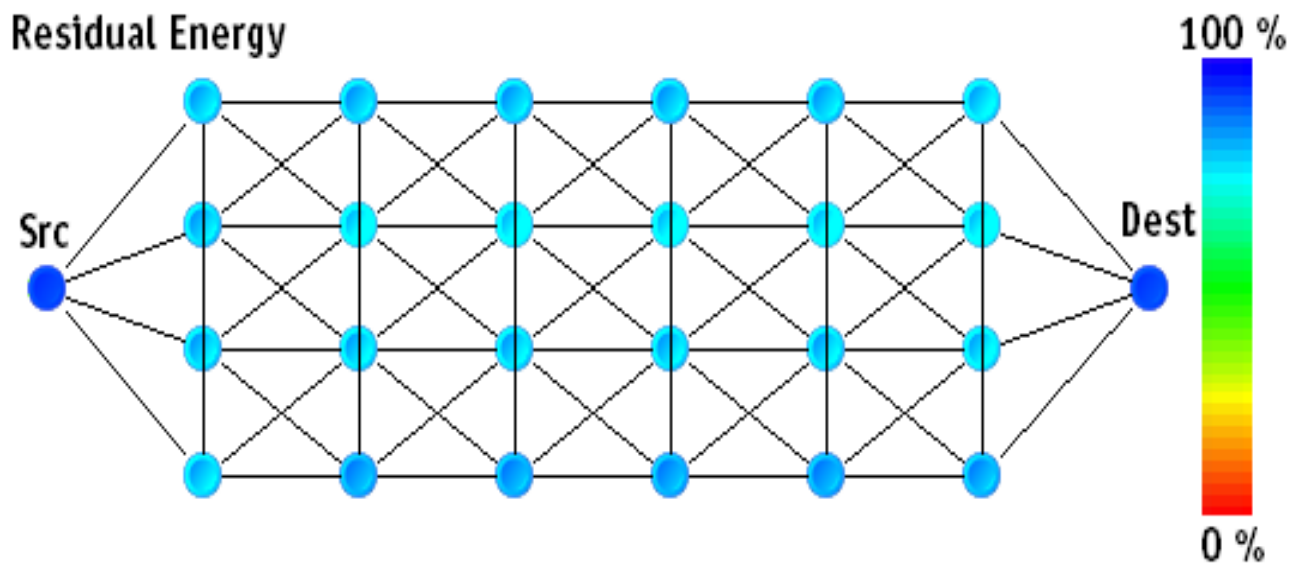

## 4.3 *Simulation Results Discussion*

### 4.3.1 *Global Energy Distribution (GED)*

The GPSR protocol always uses the closest neighbor to the destination (see GPSR behavior in a grid topology as shown in Figure 25) due to inflexible selection of the next hop node. Forwarding packets to that neighbor is costly since the distance in a greedy forwarding is considered only and longer the distance is, the most energy consuming the transmission will be. This explains why residual energy in the case of GPSR is less than in the case of AGEM as shown in figures 15 and 23.

Although the use of multiple paths in TPGF, TPGF is still more energy consuming than AGEM since it uses *"greedy"* paths.

Moreover, the energy distribution in the network is well distributed with AGEM compared to GPSR. Unlike GPSR, AGEM use various nodes to perform online multipath routing and load balancing (see Figure 26).

### 4.3.2 *Local Energy Distribution (LED)*

Figures 16-18 and 22 illustrate the average residual energy of the network partitioned in regions of 40 meters width for the plain topologies and a topology of 50 nodes with two holes. We can clearly see that the energy is uniformly consumed through the network when using AGEM routing protocol compared to GPSR and TPGF routing protocols. Moreover, AGEM uses less energy than TPGF since TPGF is a greedy routing protocol and all the explored paths use always the greedy neighbor to forward packets. The benefit of such a feature is to prevent the network from being portioned into sub networks that are completely disconnected if some nodes die because of their energy depletion.

### 4.3.3 *Packet Loss and Transmission Delay*

By using multiple paths to transmit data packets, not only the packet transmission delay has been generally reduced first by using GEAMS and AGEM as shown in figures 20 and 26, but also, this end-to-end delay has become uniform as we can see by the mean of the standard-deviation as shown in figures 21 and 27.

However, this end-to-end delay remains quite bigger than the end-to-end delay while using an offline multipath routing protocol such as TPGF. This can be explained by the fact that TPGF uses totally disjoint paths to route packets. This makes packets safe from interference problems (retransmissions).

The packet loss ratio has also been decreased as shown in figures 22 and 28 in comparison with GPSR.

The decrease in packet loss ratio and delay can be explained by the following points:

1 The use of the same path will increase the queuing delays within nodes along the routes and causes network congestion.
2 Sensor nodes have resources constraints, packet loss may occur due to the limited buffer sizes in sensor nodes.
3 In the case of topologies with holes, the perimeter routing mode employed by GPSR is not suited for burst transmissions which causes buffer over loads and packet losses.

These results demonstrate a better performance of AGEM to deliver multimedia traffic (still images in our simulation case) and provide better QoS compared to GPSR (lower the end-to-end delay and reduced packet loss ratio). AGEM is also more suitable to dense networks in which different paths to destination may exist.

## 5 Conclusion

In this paper, we have described a new algorithm namely AGEM that is suitable for transmitting multimedia streaming over WMSNs. Because nodes are often densely deployed, different paths from source nodes to the base station may exist. To meet the multimedia transmission constraints and to maximize the network lifetime, AGEM exploits the online multipath capabilities of the WSN to make load balancing among nodes. Simulation results show that AGEM is well suited for WMSNs since it ensures uniform energy consumption and meets the delay and packet loss constraint.